\documentclass{iau} 
\usepackage{graphicx}

\title[The hot bubble of BD+30$^\circ$3639]
 {Analysis of the X-ray spectrum of the hot bubble of BD+30$^\circ$3639}

\author[Detlef Sch\"onberner, Ralf Jacob, Ren\'e Heller, \& Matthias Steffen]
{Detlef Sch\"onberner, Ralf Jacob, 
  Ren\'e Heller\thanks{Present address: Max-Planck-Institut f\"ur Sonnensystemforschung, 
                        Justus-von-Liebig-Weg~3, 37077 G\"ottingen, Germany}
     \and Matthias Steffen}

\affiliation{Leibniz-Institut f\"ur Astrophysik (AIP), An der Sternwarte 16,
            14482 Potsdam, Germany
 \\ email: {\tt deschoenberner@aip.de} \\[\affilskip] }

\pubyear{2016}
\volume{323}  %% insert here IAU Symposium No.
\jname{Planetary Nebulae: Multiwavelength probes of stellar and galactic evolution}
\editors{X. Liu, L. Stanghellini \& A. Karakas, eds.}

\newcommand{\Tx}{T_{\rm X}}

\begin{document}

\maketitle

\vspace*{-1.0mm}
\begin{abstract}
 We developed a model for wind-blown bubbles with temperature
  and density profiles based on self-similar solutions 
  including thermal conduction. We constructed also heat-conduction bubbles with chemical
  discontinuities.  The X-ray emission is computed using the well-documented CHIANTI code
  (v6.0.1).   
  These bubble models are used to (re)analyse the high-resolution X-ray spectrum of the
  hot bubble of BD+30$^\circ$3639, and they appeared to be much superior to constant
  temperature approaches.

\noindent
  We found for the X-ray emission of BD+30$^\circ$3639 that 
  temperature-sensitive and abundance-sensitive line ratios computed on the basis of 
  heat-conducting wind-blown bubbles and with abundances as found in the stellar
  photosphere/wind can only be reconciled with the observations 
  if the hot bubble of BD+30$^\circ$3639 is chemically stratified, i.e. if it 
  contains also a small mass fraction ($\simeq$\,3\,\%) of 
  hydrogen-rich matter immediately behind the conduction front.  
  Neon appears to be strongly enriched, with a mass fraction of at least about 0.06. 
  \vspace{-2mm}
\end{abstract}

\vspace*{-2mm}

\vspace*{-2mm}
\section{Introduction}

  The cavities of elliptical/roundish planetary nebulae (PNe) are not empty 
  but instead filled-up
  with hot gas originating from the shock-heated fast stellar wind.  Space-born 
  X-ray telescopes (ROSAT, XMM-Newton, Chandra) revealed that these wind-blown ``bubbles''
  are the source of rather soft X-ray emission, with typical plasma temperatures 
  between 1 and 3 MK (see Kastner \etal\ 2012).    
  The low plasma temperature observed is substantially {\it below} the post-shock temperature
  of about 10 MK expected from the high observed velocities of the stellar wind.  Possible
  ``cooling'' processes are either heat conduction in the absence of magnetic fields
  (Soker 1994; Steffen \etal\ 2008)
  and/or dynamical mixing of matter across the bubble/nebula interface 
  (Toal\'a \& Arthur 2016). 
  The latest 2D simulations by \cite{TA.16} suggest that heat conduction is
  indispensable for achieving the observed high X-ray emission measures.
  
  The case of BD+30$^\circ$3639 is special in three aspects: 
  i) A hydrogen-poor wind interacts with a nebular shell of normal, hydrogen-rich composition;
  ii) it is the brightest PN X-ray source, allowing to take high-resolution spectra 
  (Yu \etal\ 2009); iii) it is a young, still unevolved nebula, obviously shortly after
  the conversion to a hydrogen-poor central star has taken place. 
  A  detailed analysis of the X-ray spectrum with respect to elemental abundance ratios and 
  the existence of a possible abundance discontinuity within the bubble produced by 
  heat conduction is thus very rewarding.
  
  All the existing spectral analyses of this nebula are  based on 
  simple plasma models with constant densities and temperatures.   
  In their analysis of the bubble of BD+30$^\circ$3639,
  \cite{yuetal.09} and \cite{nordon.09} needed a two-temperature plasma (1.9 and 3.0 MK)
  to account for the emission from species with different degree of ionisation.  
  The derived chemical abundances appeared to be rather unusual: very high ratios of
   C/O and Ne/O, exceeding by far the corresponding solar ratios. 
  
  From the physical point of view, such a low-temperature plasma is fully inconsistent with
  its outer and inner boundary conditions, viz. the nebular shell and the fast stellar 
  wind.   In the following we report briefly on our new approach to analyse the X-ray
  emission from hot bubbles by means of a physically more sound model.  Details will be
  published in a forthcoming journal paper.    %%%%% (Heller \etal, in prep.).

\vspace*{-3mm}
\section{Bubble models}

  Our newly developed analysis tool is based on self-similar solutions for wind-blown
  bubbles formed by the interaction of two spherical winds with different densities and
  velocities, as developed by %%%\cite{ZP.96, ZP.98}.
  Zhekov \& Perinotto (1996, 1998).  These so-called ZP96-bubbles contain heat conduction
  across the bubble from the reverse wind shock to the contact discontinuity/conduction front. 
  The general properties of these bubbles can be characterised as follows:\\
  {\it Inner boundary condition.}  Power-law representation of the fast stellar wind from
       an evolving 0.595 M$_\odot$ central star, time-dependent mass-loss rate and velocity,  
       adapted for BD+30$^\circ$3639 according to \cite{sandin.16}.\\
  {\it Outer boundary condition.}  Constant slow wind with 
        explored parameter ranges 
        $10^{-7} \ldots10^{-4}$ M$_\odot$\,yr$^{-1}$ and 10\ldots40 km\,s$^{-1}$. \\
  {\it Chemical composition.}  Each bubble has either a hydrogen-deficient (``WR'';
       Marcolino \etal\ 2007) or 
       hydrogen-rich (``PN'') chemical composition, with appropriate conduction coefficients. 
       Also, chemically inhomogeneous bubbles are
       constructed to allow for ``evaporated'' PN matter behind the 
       conduction front.    The WR composition is extremely helium-, carbon-, and oxygen-rich:
       He:C:O = 0.43\,:\,0.51\,:\,0.06 by mass (Marcolino \etal\ 2007).\\
  {\it X-ray emission.} The X-ray spectrum is computed by means of the well-documented
       CHIANTI code, v6.0.1 (Dere \etal\ 2009).    
       
\begin{figure}[b]
\vspace*{-2.0mm}
%%\fbox
{\includegraphics[width= 12.25cm]{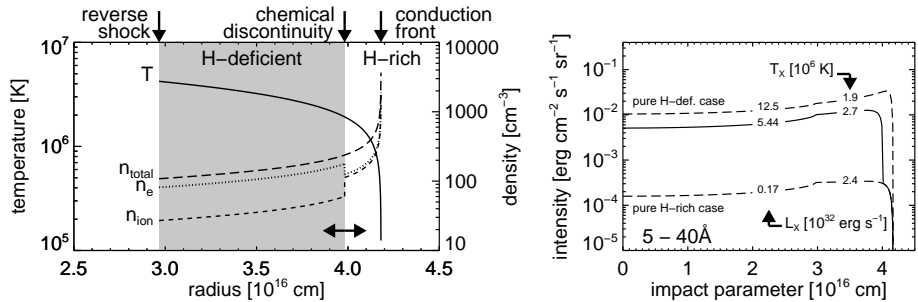} }
\vskip-2mm
\caption{\label{fig1}
         \emph{Left}:
         physical structure of a ZP96 bubble of age = 500 yr with a stratified composition
         such that ${\omega \equiv M_{\rm PN}/(M_{\rm WR} + M_{\rm PN}) = 0.03}$.  Shown
         are radial runs of electron, ion, and total particle densities, and the
         electron temperature.  %%%The central star is at $r=0$, and 
         The reverse wind shock,
         the chemical discontinuity, and the conduction front are marked by vertical arrows.
         The condition of constant pressure across the bubble leads to the density jumps
         at the chemical boundary at ${r\simeq 4{\times}10^{16}}$\,cm, or 2 MK.  
         \emph{Right}:
         the model's X-ray surface brightness integrated over 5--40 \AA.  The dashed curves
         serve for comparisons with the homogeneous WR (H-poor) and PN (H-rich)
         cases only. Remarkably is the high emission of the WR matter which is
         exclusively residing in the hotter bubble regions.          
         Thus, $\Tx$ is increased from 1.9 to 2.7 MK for the $\omega = 0.03$ case shown here, 
         while the X-ray luminosity is more than halved because of the low 
         PN-matter emissivity.
         \vspace{-2mm}
         }
\end{figure}

   Altogether, about 1000 bubbles have been generated, spanning ages from 200  to
   1000 years, corresponding characteristic X-ray temperatures, $\Tx$ (Eq.~17 in 
   Steffen \etal\ 2008),  between 1.1 and 
   3.9 MK (WR) or between 1.1 and 6.8 MK (PN).  WR-bubbles with ages of 400--500 yr
   have sizes and X-ray luminosities which correspond well to the respective  
   values observed for  BD+30$^\circ$3639.   %%%An example of 
   The structure of a bubble with a chemical discontinuity is displayed in Fig.\,\ref{fig1}.
   
%%%%%%%%%%%%%%%%%%%%%%%%%%%%%%%%%%%%%%%%%%%%%%%%%%%%%%%%%%%%%%%%%%%%%%%%%%%%%%%%%%%%
%%%%%%%%%%%%%%%%%%%%%%%%%%%%%%%%%%%%%%%%%%%%%%%%%%%%%%%%%%%%%%%%%%%%%%%%%%%%%%%%%%%%   

 \vspace*{-3mm}  

\section{Application to the case of BD+30$^\circ$3639}

\begin{figure}[h]
\hspace{-2mm}
\includegraphics[width= 13.8cm]{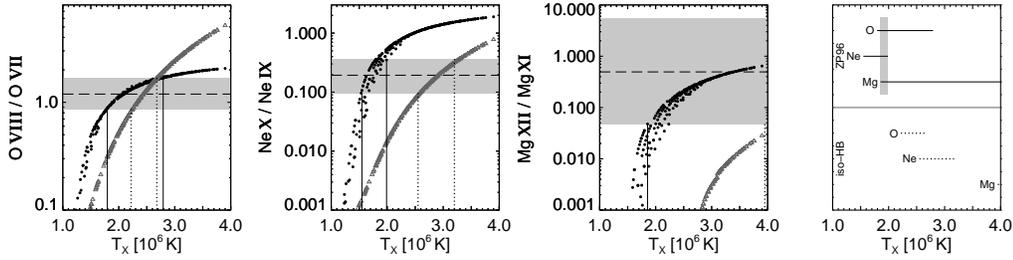}
\caption{\label{Tx}
         Line ratios for two ionisation stages of oxygen ({\it left}), neon 
         (\emph{next to left}), and magnesium ({\it next to right})
         vs.\ $\Tx$ as predicted by our ZP96 bubble models with hydrogen-deficient WR
         composition (dots) and  constant-temperature bubbles (triangles) with the same
         chemical composition.  
         The horizontal long-dashed lines give the measured values for BD+30$^\circ$3639, 
         the shaded regions indicate their uncertainties.
         The \emph{right} panel summarises the $\Tx$ determinations from the various line
         ratios and for the ZP96-bubbles and the constant-temperature approach.
%         comparison between the $\Tx$ values dereived from the ZP96 bubbles and 
%         constant-temperature plasmas.  
        }
        \vspace{-2mm}
\end{figure}

   Our analysis of the X-ray emission from BD+30$^\circ$3639's bubble rests exclusively
   on the high-resolution observations reported in \cite{yuetal.09}.
    We first used our ZP96-bubbles with WR composition to fix the characteristic plasma
   temperature $\Tx$.  The result is seen in Fig.\,\ref{Tx}.

%%%%%%%%%%%%%%%%%%%%%%%%%%%%%%%%%%%%%%%%%%%%%%%%%%%%%%%%%%%%%%%%%%%%%%%%%%%%%%%%%%%%%%%%
%%%%%%%%%%%%%%%%%%%%%%%%%%%%%%%%%%%%%%%%%%%%%%%%%%%%%%%%%%%%%%%%%%%%%%%%%%%%%%%%%%%%%%%%

   One sees that all bubbles nearly degenerate so that a unique plasma temperature can 
   easily be derived from the oxygen and neon line ratios: $\Tx =  1.9^{+0.3}_{-0.2}$ MK. 
   This temperature value, obtained from the ZP96 bubbles, is also marginally consistent with
   magnesium.  We emphasise that constant-temperature plasma models \emph{cannot} provide 
   a unique value for $\Tx$ (Fig.\,\ref{Tx}, right)!
    
\begin{figure}[h]
%\fbox
{\includegraphics*[trim= 0.3cm 10cm 10.5cm 0.5cm, width=7.9cm
                  ]{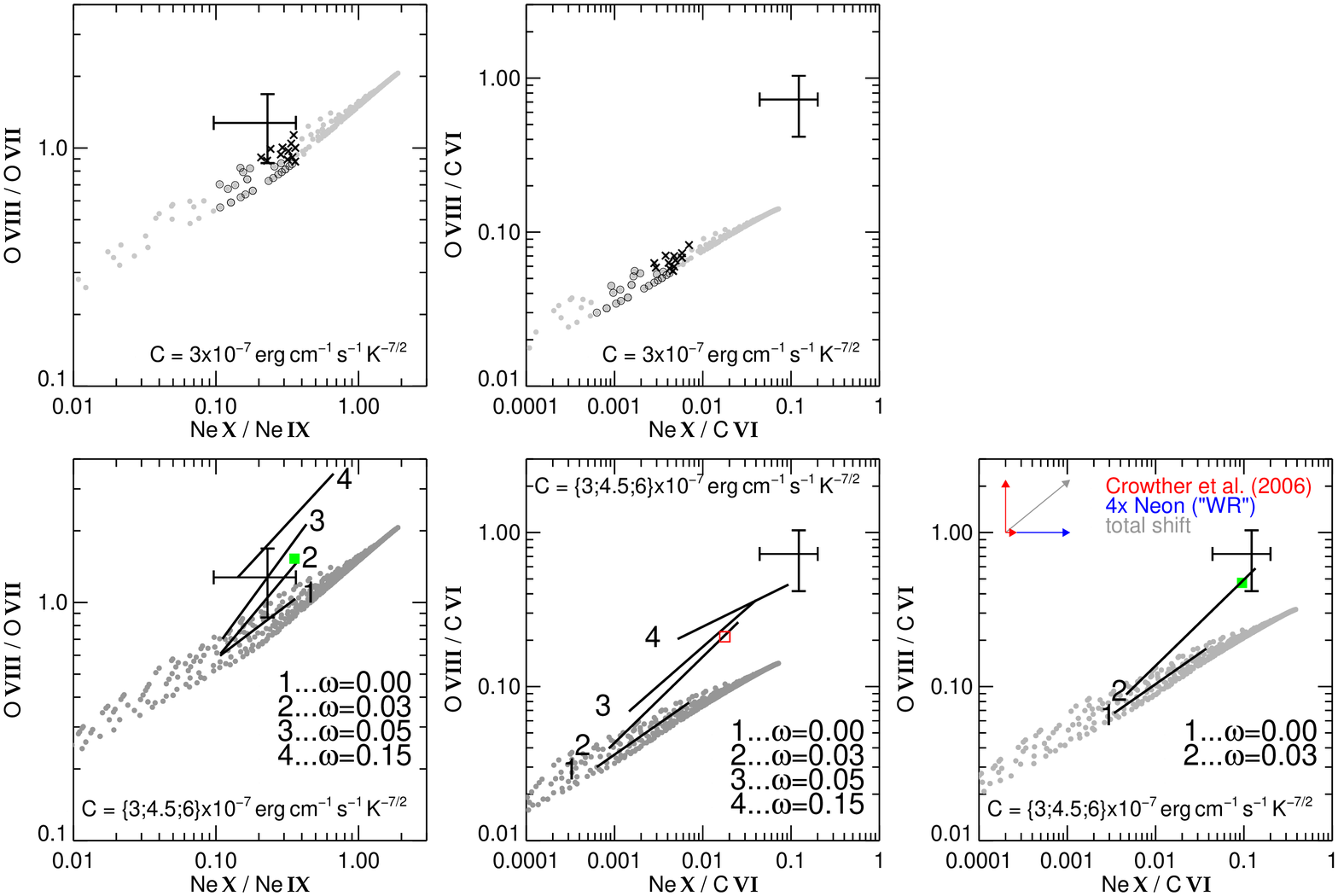} } 
\vskip-1mm
\caption{\label{consistence}
         Temperature sensitive line ratios of oxygen and neon (\emph{left}) and abundance
         sensitive line ratios (\emph{right}) as predicted by bubbles with homogeneous WR
         composition and observed values for BD+30$^\circ$3639 (big error crosses). 
         Open symbols represent bubbles which fulfil  the neon criterion, while crosses
         belong to those bubbles that fulfil additionally the oxygen temperature criterion.
         \vspace{-1mm}
         }
\end{figure}

   For the further analysis, we have to check whether those bubbles from our grid that 
   have the right temperature $\Tx$ can also reproduce the observed line ratios of carbon 
   and neon with respect to oxygen.  However, it turns out that \emph{none} of our
   bubbles with homogeneous WR composition is able to reproduce the observed line ratios
   Ne/C and O/C (Fig.\,\ref{consistence}).  The discrepancies amount to one order of
   magnitude, at least! 
    
%%%%%%%%%%%%%%%%%%%%%%%%%%%%%%%%%%%%%%%%%%%%%%%%%%%%%%%%%%%%%%%%%%%%%%%%%%%%%%%%%%%%%%%%
%%%%%%%%%%%%%%%%%%%%%%%%%%%%%%%%%%%%%%%%%%%%%%%%%%%%%%%%%%%%%%%%%%%%%%%%%%%%%%%%%%%%%%%%   

   The large discrepancy between models and observation seen in the 
   right line ratio diagram in Fig.\,\ref{consistence} cannot be remedied by changing the
   input abundances of our WR composition because they are fixed by the analyses of the
   stellar wind of BD+30$^\circ$3639 (e.g.\ Marcolino \etal\ 2007), with the exception 
   of neon whose abundance has been chosen by us.  Instead, one has to
   conclude that the bubble of BD+30$^\circ$3639 contains, as the consequence of 
   ``evaporation'',  some hydrogen-rich PN matter immediately behind the conduction front.
   The PN-matter contains much less carbon (and oxygen as well), so that even small amounts 
   of PN matter will drastically change the carbon line strengths, and to a lesser degree
   those of oxygen and neon lines because the corresponding ions reside in the inner, 
   hotter bubble regions.  

%%%%%%%%%%%%%%%%%%%%%%%%%%%%%%%%%%%%%%%%%%%%%%%%%%%%%%%%%%%%%%%%%%%%%%%%%%%%%%%%%%%%%%%%%
\begin{figure}[h]
%\fbox
{\includegraphics*[trim= 0.3cm 0.2cm 0.5cm 10.0cm, width = 12.1cm]              
                  {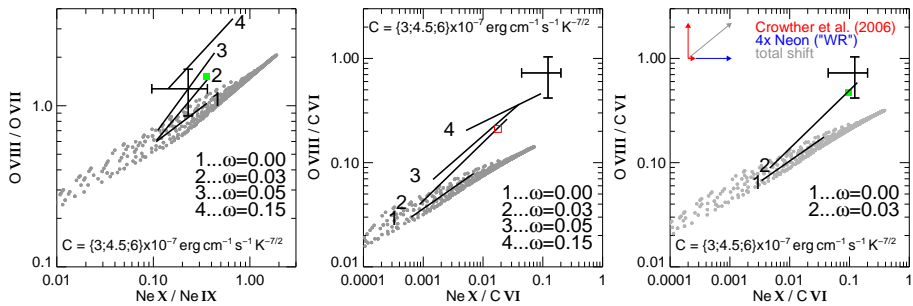}}
\caption{\label{consistence.2}
         \emph{Left and middle}:
         same as in Fig.\,\ref{consistence} but with chemically inhomogeneous bubbles 
         (black bars) added, labelled according to their $\omega$-values.  The bars
         represent the mean positions of all bubbles with the same $\omega$.
         The square seen in all panels represent our ``best-fit'' model from the 
         $\omega = 0.03$ cloud (cf.\ Fig.\,\ref{fig1}). 
         \emph{Right}:  positions of all models assuming a higher
         oxygen/carbon ratio as found by \cite{crowther.06} and the neon/oxygen ratio 
         increased by a factor of  2.5. The arrows indicate the shifts caused by the 
         respective abundance changes.
         \vspace{-1mm}
         }        
\end{figure}
%%%%%%%%%%%%%%%%%%%%%%%%%%%%%%%%%%%%%%%%%%%%%%%%%%%%%%%%%%%%%%%%%%%%%%%%%%%%%%%%%%%%%%%%%

   Figure\,\ref{consistence.2} illustrates how the inclusion of PN matter  changes the
   relevant line ratios.  The discrepancy in the abundance sensitive panel (middle) is
   substantially reduced, although there is still some disagreement between the models and the
   observations.  The square represents our ''best-fit'' model as a compromise between 
   the temperature criteria and the abundance ratios: $\omega = 0.03^{+0.02}_{-0.01}$
   with age = 500 yr (cf. Fig\,\ref{fig1}), together with appropriate bubble size and
   X-ray luminosity.

   So far we have implicitly assumed that the bubble's Ne abundance, which is not really
   constrained by photospheric/wind analyses, comes from the complete processing of
   CNO matter into N, followed by conversion to Ne during a thermal pulse on the AGB.
   This leads to a mass fraction of 0.022.  %%% which we have used here.   
   By comparing
   the observed Ne/O line ratio with the model predictions, we conclude that the Ne
   abundance in the bubble of BD+30$^\circ$3639 must be higher by a factor of about 
   2.5  (compared to oxygen), 
   i.e. we have about equal amounts (by mass) of O and Ne (at least $\simeq$\,0.06)!   
   Combined with a higher
   O/C ratio as proposed by \cite{crowther.06},  full consistency
   between both the temperature sensitive and abundance sensitive diagrams in 
   Fig.\,\ref{consistence.2} is achieved.  In this case, the Ne content would be even higher
   (up to 0.09 by mass).
   
\vspace*{-2mm}

\section{Summary and conclusion}

   We have developed a new tool for analysing the X-ray emission from wind-blown bubbles
   with thermal conduction.  Application to existing high-resolution data for
   BD+30$^\circ$3639 shows that its X-ray line emission can only be explained if the bubble
   contains
   i)  helium-, carbon-, and oxygen-enriched matter as found in the stellar photosphere/wind, 
      especially enriched by neon, and 
   ii) a shell of hydrogen-rich PN matter of about 3\,\% by mass behind the conduction
      front. 

   It appears that heat conduction is a viable option for explaining the diffuse X-ray
   emission from wind-blown bubbles.  More high-resolution X-ray observations are 
   urgently needed for further testing our analytical bubble models.  Of course, 
   new analyses of existing low-resolution X-ray spectra would also benefit from our 
   heat-conduction chemically stratified bubble models.

\end{document}